\documentclass[aps,prl,onecolumn,groupedaddress,superscriptaddress,showpacs,notitlepage]{revtex4-1}

\usepackage{type1cm}        
\usepackage{url}                             
\usepackage{makeidx}         
\usepackage{graphicx}        
\usepackage[bottom]{footmisc}

\usepackage{amsmath}
\usepackage{booktabs}
\makeindex             
\begin{document}

\title{Generating Robust and Efficient Networks Under Targeted Attacks}
  \author{V. H. P. Louzada} 
     \email{Correspondence and requests for materials should be addressed to V. H. P. Louzada (louzada@ethz.ch)}		
    \affiliation{Computational Physics, IfB, ETH Zurich, Wolfgang-Pauli-Strasse 27, 8093 Zurich, Switzerland}
  
  \author{F. Daolio}
     \affiliation{Faculty of Business and Economics, University of Lausanne, Lausanne, Switzerland}

  \author{H. J. Herrmann}
    \affiliation{Computational Physics, IfB, ETH Zurich, Wolfgang-Pauli-Strasse 27, 8093 Zurich, Switzerland}
    \affiliation{Departamento de F\'isica, Universidade Federal do Cear\'a, 60451-970 Fortaleza, Cear\'a, Brazil}

  \author{M. Tomassini}
    \affiliation{Faculty of Business and Economics, University of Lausanne, Lausanne, Switzerland}
    
\begin{abstract}
Much of our commerce and traveling depend on the efficient operation of large scale networks. Some of those, such as electric power grids, transportation systems, communication networks, and others, must maintain their efficiency even after several failures, or malicious attacks. We outline a procedure that modifies any given network to enhance its robustness, defined as the size of its largest connected component after a succession of attacks, whilst keeping a high efficiency, described in terms of the shortest paths among nodes. We also show that this generated set of networks is very similar to networks optimized for robustness in several aspects such as high assortativity and the presence of an onion-like structure.
\end{abstract}

\maketitle

\section{Introduction}
\label{sec::int}
 
In recent years, insights provided by network analysis have attracted a lot of attention from practitioners. As a result, it has been shown that several artificial (e.g. the Internet, electric-grids, etc.) and natural systems (e.g. chemical reaction networks, food networks, gene regulatory networks, etc.) present characteristics that allows one to classify them as Complex Networks. Their structure and the dynamics of phenomena taking place on them have been intensively studied~\cite{Krol2014}, thanks to the availability of large data sets~\cite{Newman2010}. 

An important aspect of a network is the capability to withstand failures and fluctuations in the functionality of its nodes and links. The design of networked infrastructures with these capabilities can be thought of as an optimization task. An early important work in this field is Albert et al.~\cite{Albert2000} where the authors showed by numerical simulations that scale-free networks, while they are robust against random removal of nodes, are much more vulnerable to the removal of nodes according to their degree. In other words, in a scale-free network if the nodes are removed in decreasing order of degree, starting with the most connected ones, then the network falls apart very quickly.

In Schneider et al.~\cite{Schneider2011}, a procedure is described that successfully modifies scale-free networks so that the largest connected component still has a considerable size after several attacks targeted at the most connected nodes. This feature guarantees that there is at least one path connecting a large number of nodes after attacks and is considered an appropriate definition of robustness\index{robustness}.
A natural question that follows is the maintenance of network efficiency\index{efficiency} after attacks, i.e., a network is efficient in this sense if
``good paths'' among nodes do not cease to exist after several targeted failures. Using a consolidated definition of 
efficiency, we propose an optimization procedure that modifies existing networks in order to 
improve their efficiency under targeted attacks.

This chapter is organized as follows. In Section~\emph{Model}, we present our measures of robustness, efficiency, and a method to optimize a specific characteristic of a network. Then, we show in Section~\emph{Results} several comparisons of optimized and unoptimized networks. We highlight the major points of our contribution in Section~\emph{Discussion}.

\section{Model}
\label{sec:2}

The proposed methodology is an extension of the work of Schneider et. al~\cite{Schneider2011}, who used a hill-climbing procedure to optimize robustness against targeted attacks. We modify this approach by adding a simulated annealing\index{simulated annealing} strategy~\cite{Kirkpatrick1983} to avoid the search getting trapped in local maxima. Previous approaches have successfully used simulated annealing to increase network\index{network} robustness~\cite{Buesser2011}. Here however we extend our focus to the following objectives: Robustness, Efficiency, and a combined measure of both. We create three sets of networks optimized for these cost functions and compare their characteristics. In what follows, we describe the cost functions and the optimization\index{optimization} procedure. 

\subsection{Robustness}
\label{robustness}

The definition of network robustness might change according to a specific application. In this work, we call an attack the removal of a node of the network, and the robustness we measure by the size of the largest connected component\index{largest connected component} (LCC) of the network after this removal, as proposed by Schneider et al.~\cite{Schneider2011}. To quantify it, we proceed with a series of attacks\index{attacks} and subsequently measure the robustness after each node removal. Hence, robustness $R$ is defined as:
\begin{align}
R = \frac{1}{N}\sum_{Q=1}^{N}S\left(\frac{Q}{N}\right)\ ,   
\end{align}
where $N$ is the number of nodes, $Q$ is the number of nodes removed from the network, and $S(q)$ is the size of the LCC after a fraction $q=Q/N$ of nodes were removed.
The attacks performed are targeted to the nodes with highest degree of the network: we find the most connected node, remove it, calculate $S(q)$, update the degrees, and find the new most connected node to repeat the process. In case two nodes have the same degree, we choose the one with the smallest index. The value $R$ is therefore unique for each network.

\subsection{Efficiency}
\label{efficiency}

One can think of network efficiency as a low cost of communication among its members. In this light, we relate efficiency with the shortest paths between all pairs of nodes, thus following Latora and Marchiori~\cite{Latora2001} who defined the network efficiency $E$ as: 
\begin{equation}
  E= \sum_{\substack{i,j=1 \\ i\neq j}}^{N}\frac{1}{l_{ij}}\ ,                                                                                                                                                                                 
\end{equation}
where $l_{ij}$ stands for the shortest path length between nodes $i$ and $j$. If $i$ and $j$ belong to separate connected components of the network, we set $l_{ij}\rightarrow \infty$ to guarantee a consistent behavior of the cost function. 

\subsection{Integral Efficiency}
\label{intefficiency}

Keeping in mind that we would like to keep the efficiency of networks after attacks, it is straightforward to modify the definition of $E$ to account for this. Hence, we define Integral Efficiency $IntE$ as:  
\begin{equation}
IntE = \frac{1}{N}\sum_{Q=1}^{N}E\left(\frac{Q}{N}\right)\ ,                                                                                                                                                                                                                                                                                                                                                                                                                                                       \end{equation}
where $E(q)$ stands for the efficiency of the network after the removal of $q=Q/N$ nodes. Nodes are removed according to a targeted attack such as in Section~\ref{robustness}. The value of $E(0)$ is the cost function $E$ defined in Section~\ref{efficiency}. By choosing this quantity instead of $E$, which does not consider nodes removal in its definition, we try to avoid that the shortest paths among nodes increases after targeted attacks. 

\subsection{Optimization procedure}
\label{optProcedure}

In their work, Schneider et al.~\cite{Schneider2011} propose a simple hill-climbing\index{hill-climbing} search to modify the network topology\index{network topology} in order to optimize the robustness $R$ whilst keeping the degree of each node fixed. This restriction in often present in the modification of artificial systems, such as electric grids where constructing a receiver for a new power line in a station might be impractical. Hence, only swaps between lines (edges in the network) are possible. A consequence of this restriction is that the underlying degree distribution of the network remains unchanged after swaps. Clearly, if we had no constraints on the degree distribution\index{degree distribution}, we could design the topology starting from scratch with the robustness and efficiency as objectives in mind, obtaining different optimal topologies. 

Next, we present an improved version of the optimization approach using simulated annealing and we describe 
it for any cost function\index{cost function} $M$ that changes after link modification\index{link modification}: 

\begin{enumerate}
\item \textbf{Initial State.} Let $G(N,E)$ be a network with $|N|$ nodes and $|E|$ edges.
 \item \textbf{Edge swap}. Choose two pairs of edges $(i,j)$ and $(k,l)$ $\in E$ randomly and
 create the network $G^*$ by deleting the edges $(i,j)$ and $(k,l)$, and adding the edges $(i,l)$ and $(k,j)$.
 \item \textbf{Acceptance probability}. Calculate the transition probability $p$ of the system as:
$$
 p = \left\{ 
  \begin{array}{l l}
   \exp \left(-\displaystyle\frac{M(G) - M(G^*)}{T}\right) & \quad \text{if $M(G^*)<M(G)$}\\
   1             & \quad \text{if $M(G^*)\geq M(G)$}\\
  \end{array} \right.
$$
 \item \textbf{Comparison}. Make $G=G^*$ with probability $p$, otherwise discard $G^*$. Return to Step 2.  
\end{enumerate}

This approach allows a network $G^*$ with $M(G^*)<M(G)$ to be chosen with finite probability. By doing this, global minima could be reached and inferior local minima could be avoided. Notice that, for the three cost functions studied here, the value of $M(G)$ is unique for each network $G$. Furthermore, by decreasing the value of $T$ according to the amount of edge swaps executed, it is possible to decrease the acceptance ratio of worst networks when an optimum point is close. We decrease the temperature as function of the number $\tau$ of edge swaps, by following the equation: $T(\tau) = 0.0001\times 0.8^\tau$. Variations to this function have shown little effect on the results. The search is stopped when a predefined amount of edge swaps\index{edge swaps} is reached.

\begin{figure}[t]
\centering
\includegraphics[width=0.8\textwidth]{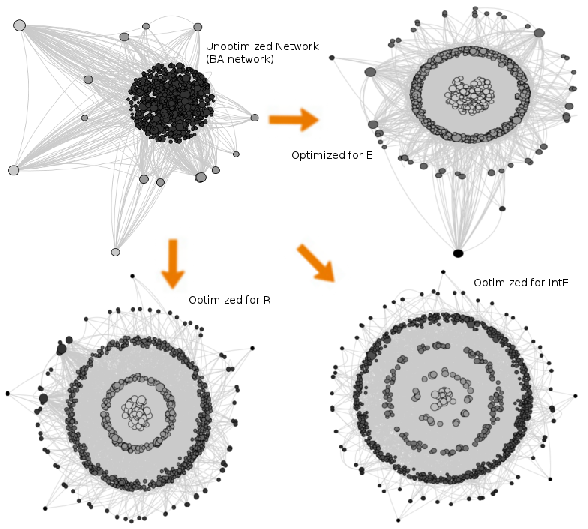}
\caption{Examples of networks belonging to each set. Networks are drawn using the k-core decomposition, represented by the different intensities of gray.}
 \label{onionsFig} 
\end{figure}

\section{Results}
\label{results}

The procedure outlined in Section~\ref{optProcedure} is applied to the cost functions: $R$ (Robustness as described in Section~\ref{robustness}), $E$ (Efficiency as described in Section~\ref{efficiency}), and $IntE$ (Integral Efficiency as described in Section~\ref{intefficiency}), starting from the same set of randomly generated of Barab\'asi-Albert (BA) networks. Hence, we created three sets of networks: \emph{Robustness set}, \emph{Efficiency set}, and \emph{Integral Efficiency set}. As a control, we compare to the original set of BA networks, from now on called the \emph{Unoptimized set}. 

The Unoptimized set is composed of $100$ networks of $n=1000$ nodes and average degree $\langle k \rangle=5.95$. The size of the networks was chosen based on a trade-off between the appearance of topological features such as the scale-free phenomenon, only present in large networks, and computational cost, as the $IntE$ cost function requires $O(n^3)$ operations to be calculated. The amount of edge swaps, $10.000$, was chosen so that for each optimized set its cost function is already statistically different from the Unoptimized set. It is possible to see that this goal was achieved by comparing the values in bold for columns $\langle E\rangle$,  $\langle R\rangle$, and $\langle IntE\rangle$ in Table~\ref{defaulttable}. To provide a visualization of the network structure created, some examples of each set are drawn in Fig.~\ref{onionsFig}.

\begin{table}[t]
\begin{center}

\caption{Average values of the cost functions, standard deviation in subscripts. Each set comprises $100$ networks with $n=1000$ nodes.  $\langle k\rangle$ = average degree, $\langle cc\rangle$ = average of clustering coefficient, $\langle r\rangle$ = average assortativity coefficient, $\langle E\rangle$ = average efficiency,  $\langle R\rangle$ = average robustness, $\langle IntE\rangle$ = average integral efficiency.}
 \label{defaulttable}

\resizebox{\textwidth}{!}{%
\begin{tabular}{lccccccc}
\toprule
Network Set & $\langle k\rangle$ &  $\langle cc\rangle$ & $\langle r\rangle$ & $\langle E\rangle$ & $\langle R\rangle$ & $\langle IntE\rangle$\\
\midrule
Unoptimized & 	$5.95_0$  & $0.0242_{0.0033}$ & $-0.085_{0.015}$ &$\mathbf{0.1486_{0.0012}}$ &$\mathbf{ 0.1837_{0.0053}}$ & $\mathbf{0.0308_{0.0011}}$\\
E & 		$5.95_0$  & $0.0053_{0.0014}$ & $-0.076_{0.011}$ & $\mathbf{0.1539_{0.0015}}$ & $0.1826_{0.0056}$ & $0.0310_{0.0012}$ \\
R & 		$5.95_0$  & $0.0200_{0.0027}$ & $0.038_{0.024}$ & $0.1459_{0.0013}$ & $\mathbf{0.2266_{0.0055}}$ & $0.0372_{0.0012}$\\
IntE & 	$5.95_0$  & $0.0195_{0.0029}$ & $0.055_{0.026}$ & $0.1456_{0.0013}$ & $0.2268_{0.0052}$  & $\mathbf{0.0391_{0.0012}}$\\
\bottomrule
\end{tabular}}
\end{center}
\end{table}

\begin{figure}[p]
\begin{center}
 \includegraphics[width=0.8\textwidth]{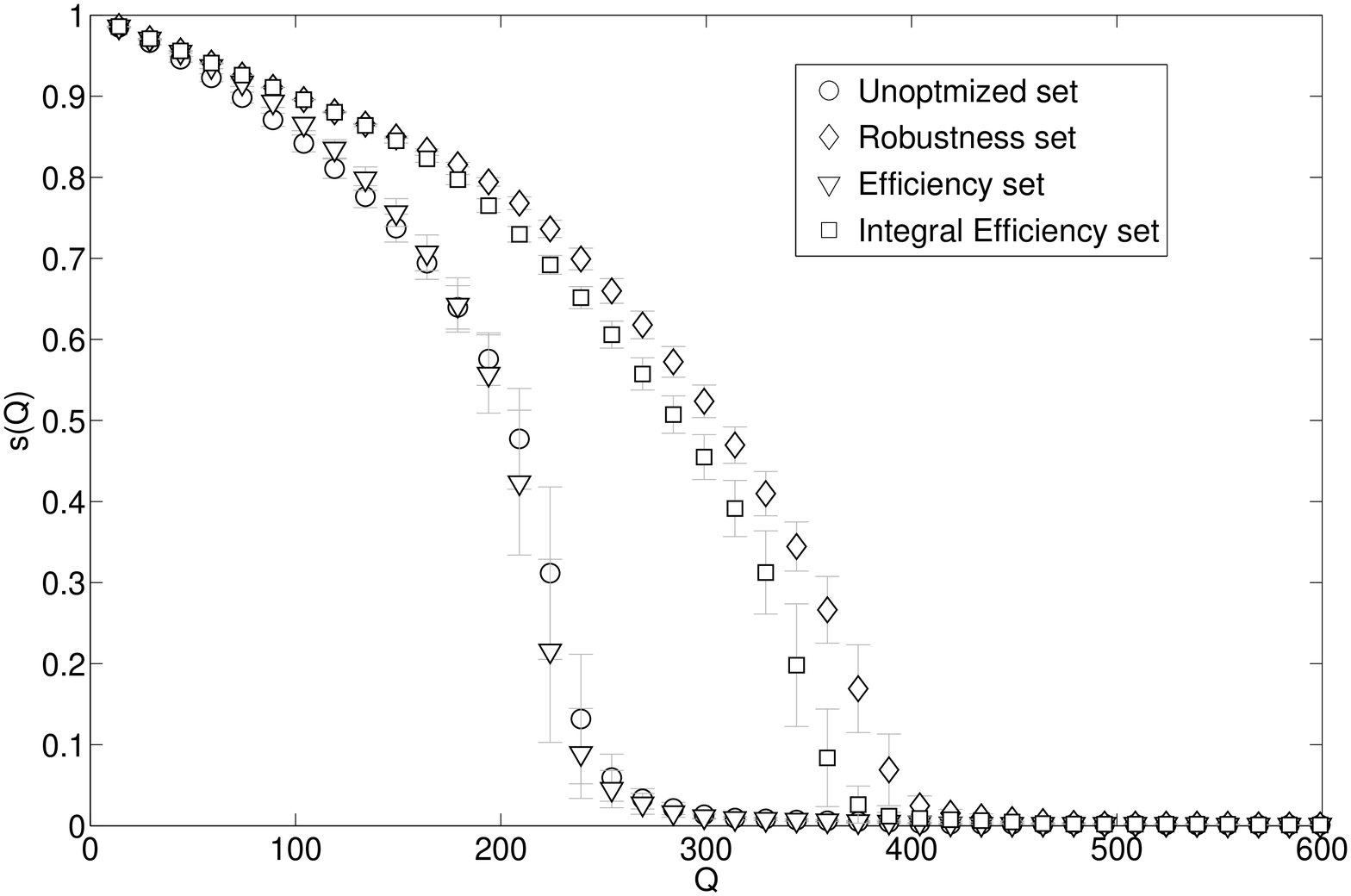}
\caption{Largest component size after the removal of $Q$ nodes. The area bellow each curve is the cost function $R$. Symbols represent sets optimized for different cost functions and are larger than the standard deviation.}
\label{plotR}
\end{center}
\end{figure}

To analyze the robustness of each set, a plot of $s(Q)$ versus $Q$ is shown in Fig.~\ref{plotR}, in which the area below each curve represents $R$ for each set. As expected, the Robustness set shows a bigger area ($23\%$ of increase), keeping a considerable size of the LCC after several attacks. Indeed, Schneider et al.~\cite{Schneider2011} obtained an improvement of almost $75\%$ for this cost function, but by using a much more exhaustive approach: their search stops after $10.000$ edge-swaps without increase in $R$. Therefore, our results show that it is possible to increase network robustness using less computational effort. The plot also shows that $E$, a cost function that does not consider attacks in its formulation, has a bad performance in this scenario. We conclude that, though more efficient, networks optimized exclusively for $E$ might not be appropriated in a realistic context, in contrast to $IntE$, which considers both effects.
Moreover, it is interesting to note also that the curves for $R$ and the Integral Efficiency set have comparable areas, considering the standard deviation of the measurements as detailed in Table~\ref{defaulttable}.

\begin{figure}[p]
\begin{center}
 \includegraphics[width=0.8\textwidth]{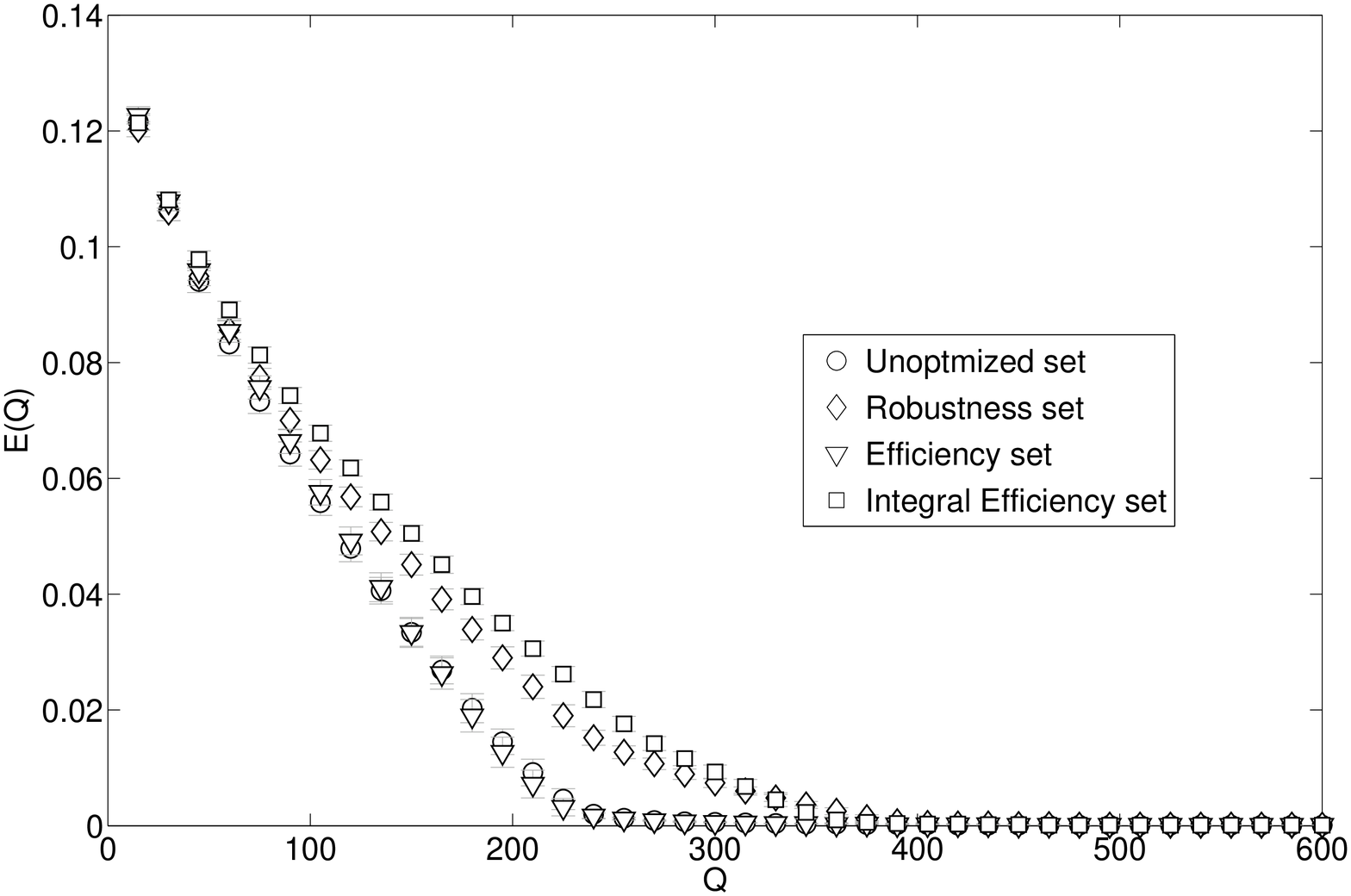}
\caption{Network efficiency $E(Q)$ after the removal of $Q$ nodes. The area bellow each curve is the cost function $IntE$. Symbols represent sets optimized for different cost functions are larger than the standard deviation.}
 \label{plotIntefficiency}
 \end{center}
\end{figure}

In Fig.~\ref{plotIntefficiency}, the cost function $IntE$ is analyzed through the plot of $E(Q)$ versus $Q$, showing that, as expected, the Integral Efficiency set has the better performance, i.e. the area under the corresponding curve is bigger. 
 Interestingly, the curve referring to the set of networks obtained by optimizing for $E$ alone shows that both have about the same performance as the unoptimized ones for this cost function (data on Column~$\langle IntE \rangle$ of Table~\ref{defaulttable}). 

Another interesting aspect of the work of Schneider et al.~\cite{Schneider2011} is the topology obtained by this optimization: a so-called onion-like structure\index{onion-like structure}. In this topology, each layer is composed of nodes connected with nodes of the same degree, with few connections between layers. A direct procedure to generate this topology can be found in the work of Wu et al.~\cite{Wu2011}.

\begin{figure}[t]
\begin{center}
 \includegraphics[width=1.0\textwidth]{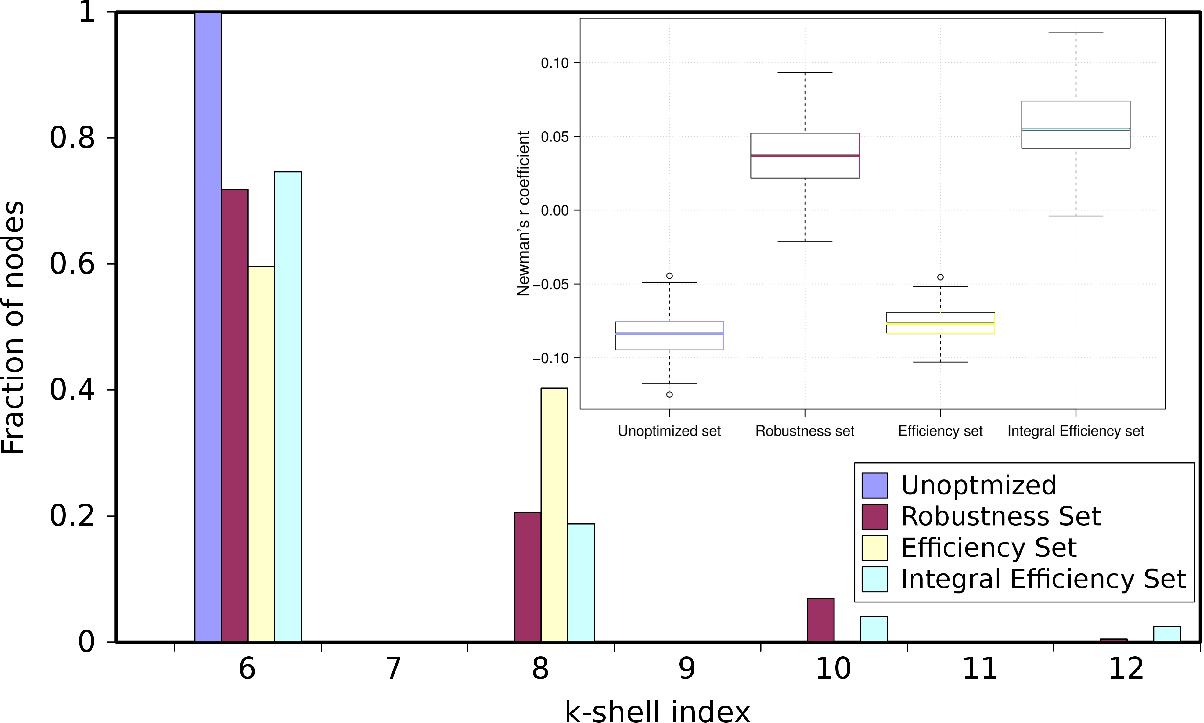}
\caption{Main plot shows the K-core decomposition for several values of $k$. It can be seen that the same network optimized for $IntE$ presents more layers than the network resulted after the optimization for $R$. Inset shows Box-and-whiskers plot of the degree assortativity through Newman's r coefficient. Thick lines depict the median value; lower and higher hinges gives the 0.25 and 0.75 quantiles, respectively; the whiskers extend to 1.5 times this inter-quantile range. Values outside this range are considered outliers and appear as circle dots in the plot.}
 \label{plotKcore}
  \end{center}
\end{figure}

To investigate the presence of an onion-like structure on our optimized sets, three quantities were analyzed. In Fig.~\ref{plotKcore}, we show the \emph{k-core} decomposition~\cite{Alvarez-Hamelin2005} for several $k$, showing that the Robustness and Integral Efficiency sets have several k-core's or layers, thus confirming a hierarchical structure of the network. The Efficiency set does not present this clear hierarchy, but has more layers than the Unoptimized set. In the inset of Fig.~\ref{plotKcore} we show that the Integral Efficiency set and the Robustness set of networks have the greater assortativity through the plot of Newman's $r$ coefficient~\cite{Newman2002}; the Efficiency set is as dissortative as the Unoptimized set. 

\begin{figure}[t]
\begin{center}
 \includegraphics[width=0.8\textwidth]{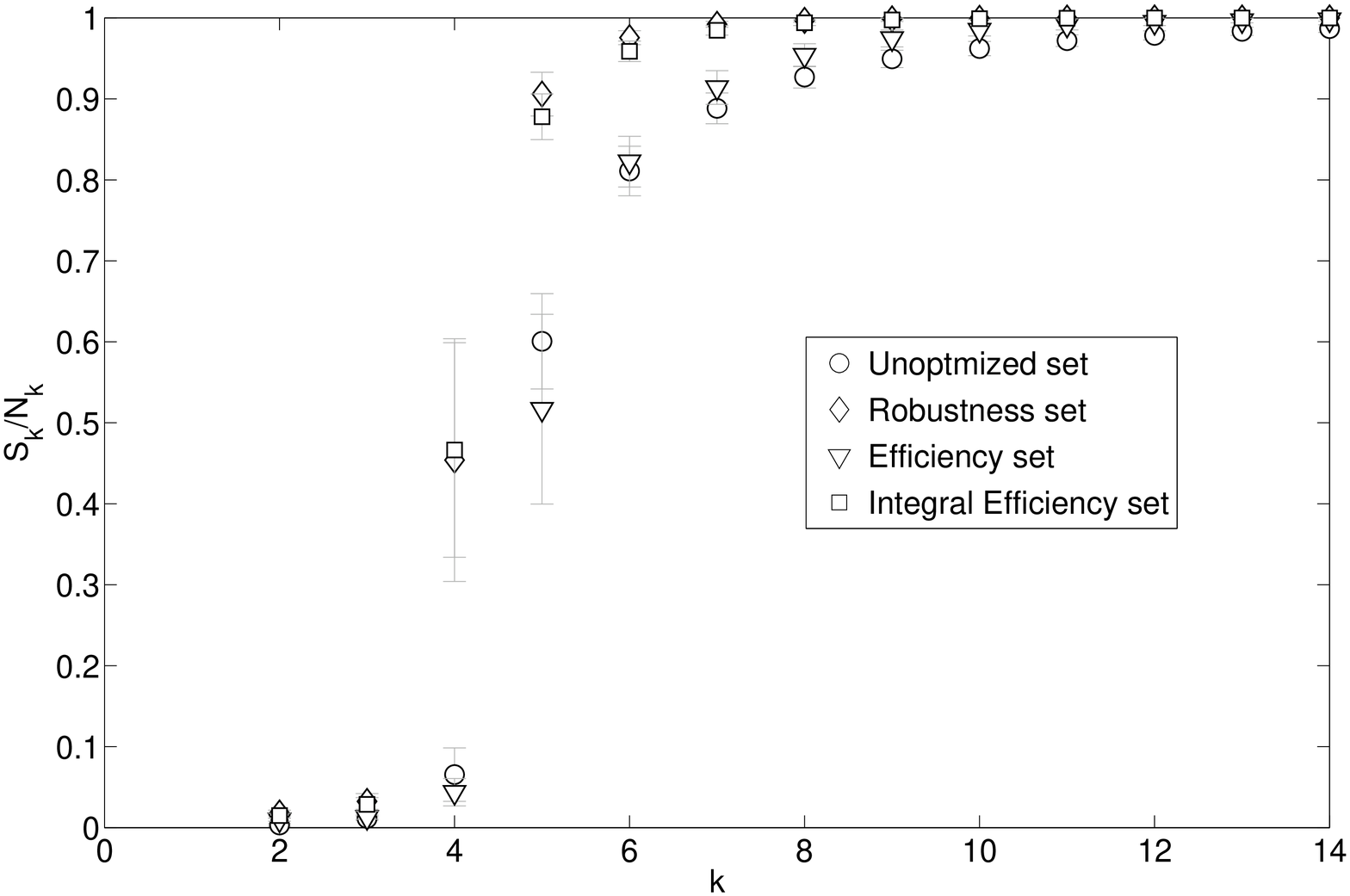}
\caption{Relative size of the largest component in networks composed of nodes of degree less than $k$. Symbols represent sets optimized for different cost functions are larger than the standard deviation.}
 \label{plotOnion}
 \end{center}
\end{figure}

Finally, we also measure the robustness for each layer of a network. To do so, we analyze the sub-graph of each network composed of $N_k$ nodes with degree smaller of equal to $k$. In this sub-graph, $S_k$ represents the size of its largest cluster. In Fig.~\ref{plotOnion}, we plot $S_k/N_k$ for several values of $k$. This plot shows that the Robustness and the Integral Efficiency sets present practically the same increase in robustness with respect to the Unoptimized set. In contrast, the Efficiency set does not show any improvement with respect to the original scale-free unoptimized networks. 

Given the several layers showed by the k-core decomposition, its dissortative nature, and the increase in robustness of each layer, we conclude that the Integral Efficiency set also has an onion-like structure similar to the Robustness set. 

\section{Discussion}

We outline here a procedure that optimizes a specific characteristic in any type of network and create three sets of BA networks with distinguishable features. Though BA networks are known to be resilient to random removals of nodes and present other interesting properties~\cite{Albert2000}, we show here a method that creates networks with a certain specific characteristic enhanced, which might be useful in some realistic scenarios.

Firstly, our results show that the Integral Efficiency set substantially improved efficiency after attacks, compared to the Robustness, Efficiency, and Unoptimized sets. Moreover, this set also sustains a large connected cluster after attacks. Therefore, this cost-function could be used to generate highly robust and efficient networks.

Another important result of our work is that networks optimized for $IntE$ also present an onion-like structure. This result suggests that this structure is generically the optimal scale-free net independently of the chosen cost function. It also helps the design of networks from scratch, as it is possible to construct scale-free networks which present this structure. 

It is also interesting to note that the Integral Efficiency set maintains several similarities with the Robustness set, such as: high assortativity, size of the largest cluster after attacks, efficiency after attacks, size of the largest cluster for each degree layer, and a hierarchical structure regarding the k-core decomposition. In fact, the Integral Efficiency set has a slightly better performance on assortativity and efficiency after attacks, while the Robustness set has a better performance on the others. 

Future works might focus on the structures of the three generated sets. The Efficiency set does not present an onion-like structure, remaining unclear if this optimization could lead to a different structure. The Integral Efficiency set might have a hidden feature that differentiates it from the Robustness set. By finding a typical structure of optimized networks, new networks could be designed from scratch with a desired feature. Also, we would like to investigate other cost functions that might lead to onion-like structures, and the case of weighted networks, as they are closer to real applications.

\begin{acknowledgments}  
This work was supported by the Swiss National Science Foundation under contract 200021 126853, the CNPq, Conselho Nacional de Desenvolvimento Cient\'i­fico e Tecnol\'ogico - Brasil, the European Research Council under grant FP7-319968-flowCSS, and ETH Zurich Risk Center.
\end{acknowledgments}

\end{document}